\documentclass[aps,prl,longbibliography,twocolumn,showpacs,noeprint,superscriptaddress ]{revtex4-2}

\usepackage{CJK}

\usepackage{graphicx}

\usepackage{amsmath}
\usepackage{dcolumn}

\usepackage{bm}
\usepackage[normalem]{ulem}
\usepackage{color}
\usepackage{hyperref}
\usepackage{siunitx}
\usepackage{epstopdf}
\newcommand{\bew}{\begin{widetext}}
\newcommand{\ew}{\end{widetext}}

\newcommand{\bp}{\mathbf{p}}

\newcommand{\br}{\mathbf{r}}
\newcommand{\bR}{\mathbf{R}}

\newcommand{\bn}{\mathbf{n}}

\newcommand{\beq}{\begin{equation}}
\newcommand{\eeq}{\end{equation}}
\newcommand{\beqn}{\begin{eqnarray}}
\newcommand{\eeqn}{\end{eqnarray}}
\newcommand{\pp}{\partial}

\allowdisplaybreaks[1]

\begin{document}
\title{ True and Quasi Long-Range Order in Malthusian Flocks}
\author{Patrick Jentsch}
\email{patrick.jentsch@embl.de}
\affiliation{Cell Biology and Biophysics Unit, European Molecular Biology Laboratory Heidelberg, Meyerhofstrasse 1, 69117 Heidelberg, Germany}
\author{Anna Erzberger}
\affiliation{Cell Biology and Biophysics Unit, European Molecular Biology Laboratory Heidelberg, Meyerhofstrasse 1, 69117 Heidelberg, Germany}
\affiliation{Department of Physics and Astronomy, Heidelberg University, 69120 Heidelberg, Germany}
\date{\today}

	\begin{abstract}
    Living matter undergoes continuous turnover. The hydrodynamic theory of Malthusian flocks describes polar active matter with turnover, but its phase diagram and nonlinear scaling behavior is not well understood. Using a nonperturbative renormalization group approach, rotationally invariant to second order in derivatives, and without defects, we explicitly obtain the strong-coupling fixed point governing true long-range order, uncover a quasi long-range ordered phase, and identify a critical point similar to, but distinct from the Berezinskii-Kosterlitz-Thouless universality class at the transition. 
	\end{abstract}

\maketitle

{\it Introduction.---}The constituents of active matter are able to utilize environmental or internal energy to drive the system out of equilibrium already at the microscopic level \cite{ramaswamy_annrev10, marchetti_rmp13}. Traditionally, the field has focused on self-propulsion and the generation of active stresses as the main microscopic out-of-equilibrium processes, which lead to a rich variety of collective behaviors such as flocking \cite{vicsek_prl95,toner_prl95,chate_rmp20,alert_annrev20,toner_b24}, active turbulence \cite{dombrowski_prl04,alert_annrev22} and motility induced phase separation \cite{fily_prl12,cates_annrev15}. 

More recently however, the role of \emph{turnover} in active matter, i.e. the production and removal of constituents, is increasingly being investigated as a distinct source of activity  that gives rise to new behaviors \cite{hallatschek_nrp23}. Particularly in living matter, such turnover is a ubiquitous and actively regulated property. For example, continuous cell division and death maintains density homeostasis in many biological tissues \cite{pellettieri_annrev07,obrien_annrev22}, including during collective cell migration in wound healing, embryonic development and cancer invasion, where cells are strongly polarized \cite{alert_annrev20,mao_natrev24}. Similarly, intrinsically polar cytoskeletal components such as actin filaments and microtubules continuously assemble and disassemble while dynamically forming large scale structures, such as spindles, asters and the cell cortex \cite{hiraiwa_prl16,goodson_csh18,banerjee_annrev20,goode_jcb23}.

Unlike self-propulsion, turnover directly breaks particle-number conservation, thereby changing the conservation laws and thus the long-wavelength physics \cite{hohenberg_rmp77,chaikin_b95, marchetti_rmp13, jentsch_a26}. 
Density fluctuations cease to be hydrodynamically relevant because the creation and decay of constituents permits the fast adaptation of the density to the remaining hydrodynamic fields.
A hydrodynamic theory of such Malthusian flocks \cite{toner_prl12} was first developed as an extension of the Toner-Tu model \cite{toner_prl95,toner_pre98,toner_pre12} for flocking almost 15 years ago. Nonetheless, its phase diagram \cite{besse_prl22, dicarlo_njp22} and universal scaling properties \cite{chate_prl24,chen_a25} are still not fully understood in two dimensions. An analytical treatment has been particularly challenging, since nonlinear effects cannot be treated perturbatively in 2D, and symmetries do not constrain the system strongly enough to fully fix the scaling behavior exactly \cite{chen_a25}; see however \cite{chate_prl24}.

To overcome this challenge, we apply here the nonperturbative renormalization group (RG) \cite{wetterich_plb93, morris_ijopa94, ellwanger_zfpc94, kopietz_b10,dupuis_pr21} to Malthusian flocks, realizing for the first time a full second-order derivative expansion (DE)\cite{balog_prl19,depolsi_pre20,delamotte_pre24} going beyond previous nonperturbative RG approaches within polar active matter \cite{jentsch_prr23,jentsch_prl24}. Due to the periodic nature of the Goldstone mode, the approach has similarities with DEs of sine-Gordon models \cite{daviet_prl19,daviet_spp22,jentsch_prd22}, however, since spatial and field rotations are intrinsically linked, the form of the DE is heavily constrained for Malthusian flocks.

Using this approach, we uncover for the first time explicitly the two-dimensional strong-coupling RG fixed point characterizing the universal scaling behavior of Malthusian flocks in the phase of true long-range order, showing that the scaling exponents predicted in Ref.~\cite{chate_prl24} are a good approximation, without the need of invoking additional symmetries beyond rotational and chiral invariance.

In addition to the Malthusian fixed point, we also found so far unpredicted features of the phase diagram: An attractive line of fixed points realizes quasi long-range order (QLRO) in Malthusian flocks, when defect fluctuations are suppressed, independently uncovered in Ref.~\cite{sezik_u26} using a field theoretic RG approach. This line has a critical end point, similar to the Berezinskii-Kosterlitz-Thouless (BKT) transition \cite{berezinskii_jetp71,berezinskii_jetp72,kosterlitz_jpc73}, yet with different scaling behavior, realizing a novel universality class \cite{hohenberg_rmp77,jentsch_a26}.

{\it Malthusian Flocks.---}
The theory of Malthusian flocks is a hydrodynamic model for collective motion, i.e., flocking, in the presence of turnover processes \cite{toner_prl12}. It can straightforwardly be obtained from the Toner-Tu equations \cite{toner_prl95,toner_pre98,toner_pre12} for ``immortal" flocks by introducing a turnover term into the equation of motion for the density. Concentrating on the state of collective motion, one readily determines that in the presence of turnover, density fluctuations $\delta \rho$ relax on a fast time scale compared to directional fluctuations due to spontaneous breaking of rotational symmetry \cite{toner_prl12}. Alternatively, but for the same reason, density fluctuations can also be discarded as a hydrodynamic variable from the very beginning such that an equation of motion (EOM) for polarity $\bp=p(\cos\theta,\sin(\theta))^T$ (equivalently velocity or momentum density) fluctuations is constructed on the basis of translational, rotational and chiral symmetry \cite{chen_a25,jentsch_a26}. Under the spin-wave approximation, i.e., when the amplitude $p$ is not allowed to fluctuate, excluding the formation of defects, the equations can further be reduced to an EOM for the directional fluctuations, i.e., the Goldstone mode only \cite{SI,toner_pre12,chen_a25},
\begin{align}
	\label{eq:eom_theta}
	\pp_t \theta &= - \lambda \cos(\theta) \pp_x \theta - \lambda \sin(\theta) \pp_y \theta + \mu_1 \nabla^2\theta + f \\
	\nonumber
	&+  \mu_2 \Big[ \cos(2\theta) ( \pp_y^2\theta - \pp_x^2\theta ) -2\sin(2\theta) \pp_x\pp_y \theta \Big] \\
	\nonumber
	&+ \mu_3 \Big[ \sin(2\theta)\Big((\pp_y \theta)^2-(\pp_x \theta)^2\Big) + 2\cos(2\theta) \pp_x\theta \pp_y \theta\Big] \ .
\end{align}
where the noise term $f$ is a zero-mean Gaussian noise with statistics,
\begin{equation}
	\langle f(t,\br) f(t^\prime,\br^\prime) \rangle = 2 D \delta^{2}(\br-\br^\prime)\delta(t-t^\prime) \ .
\end{equation}
Ref.~\cite{sezik_u26} arrives at the same equation \eqref{eq:eom_theta} using a different construction.
Due to rotational,
\begin{equation}
    \theta(t,\br)\to \theta(t,\br') + \psi\ ,\ \br\to\br' = \bR^{-1}(\psi)\cdot\br \ ,
\end{equation}
and chiral symmetry
\begin{equation}
    \theta(t,\br)\to -\theta(t,\br')  ,\ x\to x' = x ,\ y\to y' = -y \ ,
\end{equation}
where $\bR(\psi)$ is the associated two-dimensional rotation matrix,
the form of the EOM is drastically constrained. At second order in derivatives, yet with full nonperturbative $\theta$-dependence, the EOM is fully characterized by five parameters only, self-advection $\lambda$, (nonlinear) isotropic and anisotropic orientational diffusion $\mu_1$, $\mu_2$, $\mu_3$, and noise amplitude $D$. This structure is conserved under the RG.

{\it Functional Renormalization Group Approach.---}
We now analyze the EOM of the Goldstone mode \eqref{eq:eom_theta} using the Functional Renormalization Group Approach \cite{canet_jopa11,dupuis_pr21,SI} based on the Wetterich equation \cite{wetterich_plb93,ellwanger_zfpc94,morris_ijopa94},
\begin{equation}
	\label{eq:wetterich}
	\pp_k \Gamma_k =\frac{1}{2} {\rm Tr} \left[ \left(\Gamma^{(2)}_k +R_k\right)^{-1} \pp_k R_k\right]\ ,
\end{equation}
which is an exact renormalization group equation for the scale-dependent effective average action $\Gamma_k$, that can be understood as the effective theory at an inverse length scale $k$. $\Gamma_k^{(2)}$ is the second order functional derivative of $\Gamma_k$, $R_k$ is a regulator term that freezes out fluctuations at scales larger than $k^{-1}$ (for more details see \cite{SI}) and ${\rm Tr}$ sums over all degrees of freedom. To apply this approach to nonequilibrium systems defined by a Langevin equation like Eq.~\eqref{eq:eom_theta}, it must first be mapped to an action $S$ using the Martin-Siggia-Rose-de Dominicis-Janssen (MSRDJ) formalism \cite{martin_pra73,dedominicis_jpc76,janssen_zpb76,canet_jopa11}, which serves as an initial condition for $\Gamma_k$:
\begin{widetext}
\begin{align}
	\label{eq:action}
	\nonumber
	S[\bar \theta,\theta] &= \int_{\tilde \br} \Bigg\{-D\bar\theta^2 +\bar \theta \Bigg(\gamma\partial_t\theta+ \lambda \cos(\theta) \pp_x \theta + \lambda \sin(\theta) \pp_y \theta - \mu_1 \nabla^2\theta \\
	&-  \mu_2 \Big[\cos(2\theta) ( \pp_y^2\theta - \pp_x^2\theta ) -2\sin(2\theta) \pp_x\pp_y \theta \Big] - \mu_3 \Big[ \sin(2\theta)\Big((\pp_y \theta)^2-(\pp_x \theta)^2\Big) + 2\cos(2\theta) \pp_x\theta \pp_y \theta\Big] \Bigg) \Bigg\}\ ,
\end{align}
\end{widetext}
where $\int_{\tilde \br}= \int d^d\br d t$, $\gamma=1$ was introduced to allow for renormalization of the time derivative term and $\bar \theta$ is the response field introduced by the formalism. Eq.~\eqref{eq:wetterich} then describes how the effective theory changes upon the removal of a regulator term $R_k$, such that at $k=0$, all fluctuations are included.

Choosing $R_\Lambda\sim\Lambda^2$ and $R_0=0$ ensures the boundary conditions $\Gamma_\Lambda=S$, and $\Gamma_k=\Gamma$, i.e, the full effective average action. 
Additionally, out of equilibrium $R_k$ must be chosen such that causality is preserved \cite{SI,canet_jopa11}.

To define our approach, we have to specify an ansatz for $\Gamma_k$ and a regulator. Since $S$ was constructed from Eq.~\eqref{eq:eom_theta}, taking $\Gamma_k$ of the same form as Eq.~\eqref{eq:action}, albeit with $k$-dependent couplings, amounts to a full second-order derivative expansion of $\theta$. The noise is expanded to zeroth order in derivatives. Since the zero mode of the noise is not renormalized \cite{toner_prl12,chate_prl24,chen_a25}, it is further exact to all orders in $\bar \theta$ to this order in derivatives. As a regulator, we choose a simple mass-like cutoff,
\begin{equation}
	R_k(q) = \mu_y k^2 \begin{pmatrix}
		0 & 1 \\
		1 & 0
	\end{pmatrix} \ ,
\end{equation}
where $\mu_y = \mu_1+\mu_2$ (equivalently we define $\mu_x=\mu_1-\mu_2$), which allows for analytic evaluation of the flow equation. We have also investigated a sharp cutoff showing that our results for the fixed point structure and scaling exponents do not depend on this choice \cite{SI}.

\begin{figure*}
    \centering
    $
    \begin{array}{cccc}
         \includegraphics[width=0.5\linewidth]{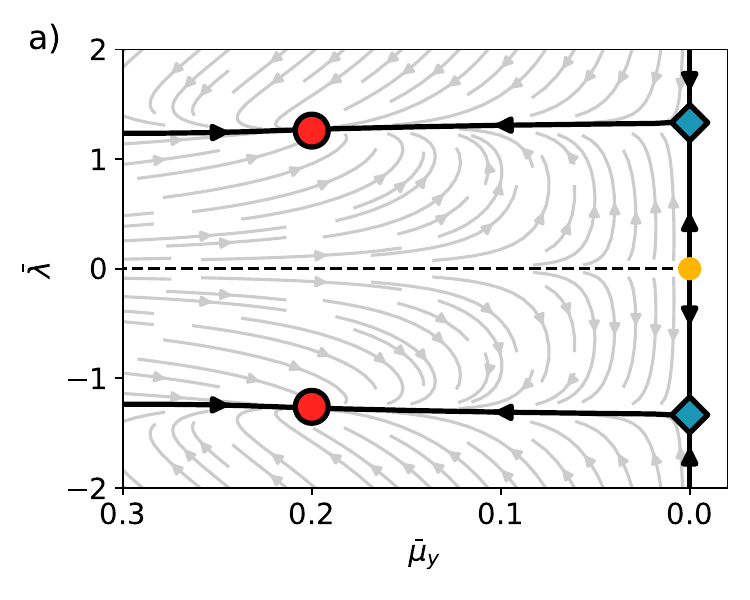} &  \includegraphics[width=0.5\linewidth]{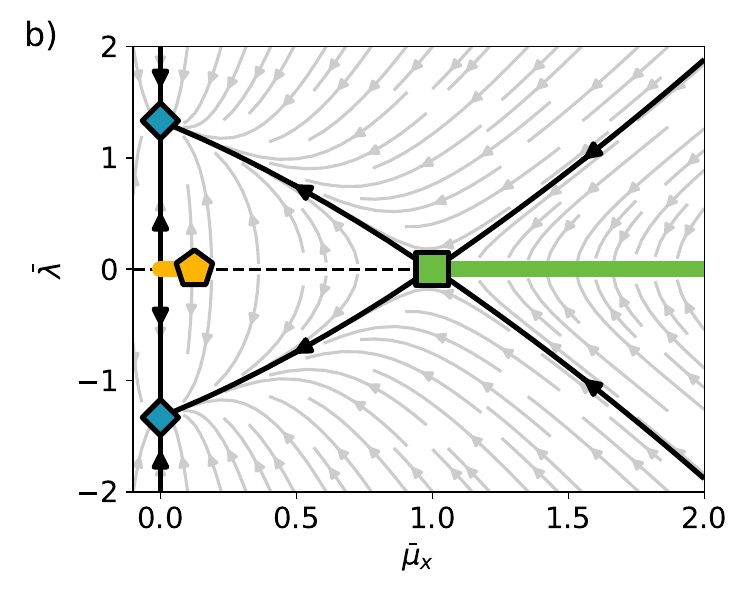}
    \end{array}$
    \caption{Nonperturbative Renormalization group flow for $\bar\mu_x=0$ and $\bar\mu_3$ set to the solution given by the remaining couplings as detailed in \cite{SI} (a), and for $\bar\mu_3=0$ and $\bar\mu_x=\bar\mu_y$ in (b). We obtain the Malthusian universality class (red circle), the universality class originally proposed by Toner \cite{toner_prl12} (blue diamond) and the attractive line of fixed points (green line), ending in a critical point (green square). For comparison with the equilibrium BKT universality class, its line of fixed points (yellow line), which is stable against defect fluctuations, and critical end point, i.e., the critical BKT transition (yellow pentagon) are marked. In particular, the BKT line of fixed points is unstable against the active self-propulsion $\lambda$, whereas the Malthusian line of fixed points is likely unstable against defect fluctuations. }
    \label{fig:placeholder}
\end{figure*}
{\it Results.---} The RG equations can now be deduced from Eq.~\eqref{eq:wetterich}, 
\begin{align}
	\pp_\ell \bar \mu_x  &= \frac{1}{2}(-3\eta_x-\eta_y)\bar \mu_x +  \eta_x \bar \mu_x \ , \\
	\pp_\ell \bar \mu_y  &= \frac{1}{2}(-3\eta_x-\eta_y)\bar \mu_y +  \eta_y \bar \mu_y \ , \\
    \label{eq:flow_lambda}
	\pp_\ell \bar \lambda  &= \Big(1-\frac{\eta_x+5\eta_y}{4}\Big)\bar \lambda +  \frac{2-\eta_y}{4}\frac{\bar\mu_3-\bar\mu_x-\bar\mu_y}{\bar \mu_y} \bar \mu_x \bar \lambda \ , \\
	\pp_\ell \bar \mu_3  &= \frac{1}{2}(-3\eta_x-\eta_y)\bar \mu_3 \\ \nonumber  &\hspace{-0.5cm}+ \frac{2-\eta_y}{96} \Big(\bar \lambda^2(13\bar\mu_3-13\bar \mu_x-23\bar\mu_y)-192\bar\mu_3\bar\mu_x\Big) \ , 
\end{align}
where $\ell=-\log k/\Lambda$ is the ``RG time" and $\eta_x$ and $\eta_y$ are the graphical corrections of $\mu_x$ and $\mu_y$ which characterize their anomalous scaling at a fixed point, see \cite{SI} for details. $D$ and $\gamma$ are not renormalized. %In particular, projecting the flow equations in the comoving frame is a vital ingredient. 
To obtain fixed point solutions, the flow equations are expressed in dimensionless units,
\begin{align}
    \nonumber
	\bar \mu_x &= \frac{D}{4\pi \gamma} \frac{1}{\sqrt{\mu_x^3 \mu_y}} \mu_x \ , &
	\bar \mu_y &= \frac{D}{4\pi \gamma} \frac{1}{\sqrt{\mu_x^3 \mu_y}} \mu_y \ , \\
	\bar \lambda &= \sqrt\frac{D}{{4\pi\gamma k^2}\sqrt{\mu_x\mu_y^5}} \lambda \ , &
	\bar \mu_3 &= \frac{D}{4\pi \gamma} \frac{1}{\sqrt{\mu_x^3 \mu_y}} \mu_3 \ ,
\end{align}
whose form follows from requiring the anomalous dimensions $\eta_{x/y}=\partial_l\mu_{x/y}/\mu_{x/y}$ to be dimensionless. Note that $\bar\mu_x$ is proportional to the amplitude of the correlation function, and thus has the same scaling dimension $\bar\mu_x\sim |\br|^{2\chi}$. For true long-ranged order $\chi<0$ we can thus anticipate that $\bar\mu_x$ vanishes at the Malthusian fixed point.

The RG flow, together with the locations of the fixed points is visualized in Fig.~\ref{fig:placeholder}. We find indeed the fixed point corresponding to the universality class of Malthusian flocks ($\bar\lambda^*=\sqrt{8/5}$, $\bar\mu^*_y=-\bar\mu_3^*=1/5$, $\bar\mu_x^*=0$), the fixed point originally proposed by Toner in Ref.~\cite{toner_prl12} ($\bar\lambda^*=4/3$, $\bar\mu^*_y=\bar\mu_x^*=\bar\mu_3^*=0$) and, surprisingly, an attractive line of fixed points with a critical end-point ($\bar\lambda^*=0$, $\bar\mu^*_y=\bar\mu_x^*=1$, $\bar\mu_3^*=0$).

{\it Quasi long-range order and novel critical point.---} The attractive line of fixed points implies the existence of QLRO in the Malthusian flocking model. If the initial amplitude of fluctuations $\bar\mu_x=D/4\pi\gamma\sqrt{\mu_x\mu_y}$ is large enough (corresponding to the limit of large noise or weak diffusion) while nonlinearities are weak enough, the RG flow is attracted to the free, Gaussian theory. In particular, here $\bar\mu_x=\bar\mu_y$ since otherwise the nonlinearity $\bar\mu_2 =(\bar\mu_y-\bar\mu_x)/2$ would have to have a nonzero fixed point value. 

A standard result of 2D Gaussian theories is that the logarithmic divergence of the correlation function \cite{cardy_b96,kardar_b07},
\begin{equation}
\label{eq:corr}
\frac{1}{2}\langle [\theta(0,\br)-\theta(0,0)]^2 \rangle 
\to \frac{D}{2\pi \gamma \mu_1} \log \Lambda r =  2\bar\mu_x^* \log\Lambda r  \ ,
\end{equation}
where $\bar\mu_x^*$ is the fixed point value taken once the RG flow has converged to the attractive line, implies the nonuniversal scaling law,
\begin{equation}
\label{eq:nontriv_scaling}
     \langle [\bn(0,\br)-\bn(0,0)]^2 \rangle = e^{-\frac{1}{2} \langle [\theta(0,\br)-\theta(0,0)]^2 \rangle } = (\Lambda r)^{-2\bar\mu_x^*} \ ,
\end{equation}
with a power law that depends on where exactly on the line the RG flow has converged to. At the critcal end-point $\bar\mu_x^*=\eta_c^{\rm Malth.}/2=1$, however, the exponent becomes universal, closely ressembling the behavior at the Berezinskii-Kosterlitz-Thouless (BKT) transition \cite{berezinskii_jetp71,berezinskii_jetp72,kosterlitz_jpc73}, but with a distinct critical exponent.  Our results can also be understood perturbatively. As pointed out in Ref.~\cite{sezik_u26}, near the free theory $\cos\theta$ and $\sin\theta$ have a scaling dimension $-\bar\mu_x$ due to Eq.~\eqref{eq:nontriv_scaling}, implying that $\lambda$ has scaling dimension $1-\bar\mu_x$ if the full nonlinear structure of the nonlinearity is accounted for. Large noise or weak diffusion can therefore make $\lambda$ irrelevant with a transition occuring at $\bar\mu_x=1$. Near the free theory, our flow equations agree with the perturbative result. The logarithmic corrections to scaling at the critical point are also determined in Ref. \cite{sezik_u26}. 

The analysis presented here, however, crucially hinges on the initial assumption, made in the derivation of Eq.~\eqref{eq:eom_theta}, that amplitude fluctuations can be neglected, implicitly assuming $\theta$ is a smooth field, i.e., the spin-wave approximation. In equilibrium, it is well-known that this assumption breaks down in the limit of large fluctuations. Nonperturbative amplitude fluctuations, allow the creation of defect pairs which can eventually unbind and destroy QLRO, which is precisely what is described by the BKT transition. While we are unable to derive a theory that accurately captures the nonperturbative effect of defect pairs, we can nevertheless apply standard results to the equilibrium sector of our theory, i.e., at $\lambda=\mu_2=\mu_3=0$, which coincides with the equilibrium $O(2)$ model with model A dynamics in the spin-wave approximation. In particular, the attractive line of fixed points, lies entirely within this sector.

Identifying $\bar\mu_x=1/4\pi K$, where $K$ is the stiffness parameter of the XY-model \cite{cardy_b96,kardar_b07}, we can overlay the attractive line and critical end-point of the BKT transition at $\bar\mu_x^*=\eta_c^{\rm BKT}/2=1/8$ onto our flow diagram, Fig.~\ref{fig:placeholder}. While our RG calculation predicts stable QLRO at large noise amplitude, the inverse is true for the BKT transition. Further, the BKT transition describes a phase transition from QLRO to true disorder. The critical point uncovered here however, describes a transition from QLRO to true long-range order. Using the equilibrium result, there is no overlap between the two lines of fixed points, suggesting that QLRO is generically destablized in Malthusian flocks, either by activity or defect fluctuations. As discussed below, it is however not clear whether this equilibrium result applies in general to Malthusian flocks.

{\it The Malthusian universality class.---} Our nonperturbative RG approach allows us to explicitly obtain the universality class of Malthusian flocks \cite{chate_prl24,chen_a23}. We find, at the level of approximation of our approach, scaling exponents that agree with the exponents obtained in \cite{chate_prl24}, i.e., $\eta_x=1/4$, $\eta_y=3/4$, and therefore,
\begin{align}
    \chi=&-(\eta_x+\eta_y)/4&=&-1/4 \ , \\
    \zeta=&1-(\eta_y-\eta_x)/2&=&\ \,\,3/4 \ , \\
    z=&2-\eta_y &=&\ \,\, 5/4 \ .
\end{align}
Unlike Ref.~\cite{chate_prl24} we do not need to invoke any additional symmetry other than rotational symmetry to obtain this result. Instead we show that the coupling $\lambda$ is generically renormalized \eqref{eq:flow_lambda}. As the Malthusian fixed point is approached, this graphical correction vanishes asymptotically. Relating our flow equations to the dynamic RG framework \cite{forster_pra77}, we show that at the one-loop level the graphical correction to $\lambda$ couples only to irrelevant nonlinearities when rotational symmetry is respected \cite{SI}. There, we also show that the same argument used in Ref.~\cite{chen_a25} to determine that the nonlinearity $\theta^2\pp_x\theta$ is relevant can be applied to two additional nonlinear terms stemming from $\mu_2$.

{\it Discussion \& Outlook.---}
Performing a nonperturbative RG analysis of 2D Malthusian flocks, fully respecting rotational symmetry at second order in derivatives, we explicitly determined the RG fixed point and scaling exponents of its ordered phase. At the level of our approximation, the exponents agree with the prediction in Ref.~\cite{chate_prl24}, however, without invoking additional symmetries \cite{chen_a25}. While our nonperturbative approach does not produce exact results, it nevertheless shows that these scaling exponents are an excellent approximation and that corrections due to higher order terms capturing e.g., two-loop effects, are likely small.

Surprisingly, we also find an attractive line of fixed points where the self-advection $\lambda$, characterizing active self-propulsion, becomes irrelevant, realizing a phase of quasi long-range order (QLRO). A critical end point, similar to but distinct from the Berezinskii-Kosterlitz-Thouless transition \cite{berezinskii_jetp71,berezinskii_jetp72,kosterlitz_jpc73}, marks the transition from equilibrium quasi long-range to active true long-range order, realizing a novel universality class. This behavior was uncovered independently in Ref.~\cite{sezik_u26} using perturbative field-theoretic RG. Complementarily to our work, the authors elucidate the logarithmic corrections to scaling at the critical point. 

Though not captured in our approach, we expect defect fluctuations to destabilize the QLRO phase due to the corresponding behavior of the equilibrium XY model. Ref.~\cite{besse_prl22} suggests that such fluctuations might even generically destabilize the phase of true long-range order. These genuinely nonperturbative effects can be treated either through the inclusion of amplitude fluctuations \cite{jakubczyk_pre14,defenu_prb17}, dual mappings \cite{villain_jphysf75,sieberer_prb16} or explicit introduction of defect pairs \cite{kosterlitz_jpc73,shankar_prl18,juelicher_pre22}. Nevertheless, it has been shown that in several active systems nonequilibrium effects can modify the critical exponent of the defect unbinding transition to non-universal values of $\eta_c^{\rm BKT}\neq1/4$ \cite{shankar_prl18,juelicher_pre22,shi_prl23,carenza_prl25}. In more complex systems including living matter, additional processes might thus modulate or suppress defect unbinding \cite{hirota_pre26}, such that $\eta_c^{\rm BKT}>2>\eta_c^{\rm Malth.}$, raising the question whether biological systems are thereby able to access these novel regimes of collective organisation which could provide functional advantages.

In principle, the divergent susceptibility of the QLRO state for example allows for the long-range propagation of stimuli through the orientation field, enabling adaptive behaviors such as collective reorientation in response to weak environmental influences even when globally disordered. Moreover, that the transition to true long-range order is controlled by the self-advection $\lambda$ alone suggests that a collectively migrating state can be entered by simply increasing motility without the need to change alignment interactions. The nonlinear response of such a state is then governed by the scaling exponents obtained in this work.

More generally since our conclusions rely only on fundamental symmetries and conservation laws, our results are expected to be relevant for any active polar system at density homeostasis.

{\it Acknowledgments.---}We thank Emir Sezik, Gunnar Pruessner, Ananyo Maitra, Chiu Fan Lee, John Toner, Sriram Ramaswamy and Leiming Chen for useful discussions and critical reading of our manuscript. PJ was supported by the EMBL Interdisciplinary Postdoctoral Fellowship (EIPOD-LinC) program.

\bibliography{references}

\end{document}

\title{ Supplemental Material to:\\True and Quasi Long-Range Order in Malthusian Flocks}
\author{Patrick Jentsch}
\email{patrick.jentsch@embl.de}
\affiliation{Cell Biology and Biophysics Unit, European Molecular Biology Laboratory Heidelberg, Meyerhofstrasse 1, 69117 Heidelberg, Germany}
\author{Anna Erzberger}
\affiliation{Cell Biology and Biophysics Unit, European Molecular Biology Laboratory Heidelberg, Meyerhofstrasse 1, 69117 Heidelberg, Germany}
\affiliation{Department of Physics and Astronomy, Heidelberg University, 69120 Heidelberg, Germany}
\date{\today}

\maketitle

\section{Equation of Motion}

In the hydrodynamic limit, density fluctuations cease to be relevant and the polar order parameter $\bp$ is the only relevant hydrodynamic field \cite{toner_prl12,chate_prl24,chen_a25}. From rotational, translational and chiral symmetry, one can immediately infer the EOM for Malthusian flocks,
\begin{align}
    \nonumber
	\label{eq:eom}
	\pp_t p_i &+ \lambda_1 p_j \pp_j p_i +\lambda_2 p_i  \pp_j p_j + \lambda_3 p_j \pp_i p_j = f_{p,i} -U p_i +\mu_A\pp_j\pp_j p_i + \mu_B \pp_i \pp_j p_j\\
    \nonumber
    & + \mu_C p_j p_k \pp_j\pp_k p_i + \mu_D p_i p_j\pp_k\pp_k p_j + \mu_E p_i p_j\pp_j\pp_k p_k + \mu_F p_jp_k \pp_j\pp_i p_k\\
    \nonumber
    & + \mu_G p_i (\pp_j p_k) (\pp_j p_k) + \mu_H p_i (\pp_j p_j) (\pp_k p_k)+ \mu_I p_i (\pp_j p_k) (\pp_k p_j)  \\
    \nonumber
    & + \mu_J p_j (\pp_j p_i) (\pp_k p_k) + \mu_K p_j (\pp_j p_k) (\pp_i p_k) + \mu_L p_j (\pp_j p_k) (\pp_k p_i) \\
    & + \mu_M p_j (\pp_i p_j) (\pp_k p_k) + \mu_N p_j (\pp_k p_j) (\pp_k p_i)  + \mu_O p_j (\pp_k p_j) (\pp_i p_k) 
\end{align}
where all couplings, i.e., the $\lambda$'s, $U$ and the $\mu$'s may arbitrarily depend on $|\bp|$, and we thus have considered the most general form of the EOM for {\it arbitrary} order in $\bp$ and second order in derivatives. The noise term $\bff_p$ is a Gaussian noise with zero mean and statistics,
\begin{equation}
	\langle f_{p,i}(\tilde \br) f_{p,j}(\tilde \br^\prime) \rangle = 2D \delta_{ij} \, \delta^{d+1}(\tilde \br-\tilde \br^\prime)\ ,
\end{equation}
where $\tilde \br = (t,\br)$ and $\delta^{d+1}$ is the $d$-dimensional Dirac delta function and $D$ characterizes the amplitude of fluctuations. 

If, microscopically, particles prefer to align their direction of motion, the force term $U$ typically has the form,
\begin{equation}
	U(|\bp|) > 0 \ \ \ \text{if} \ \ \  |\bp| > p_0  \ \ \ \text{and} \ \ \ U(|\bp|) < 0 \ \ \ \text{if} \ \ \ |\bp| < p_0 \ ,
\end{equation}
for some value $p_0$. Mean field theory thus predicts that the system is in an ordered phase with $\langle \bp \rangle = p_0 \hat \bx $. The average direction is chosen spontaneously through fluctuations and initial conditions. Without loss of generality, we label this direction as the $x$-direction. Hats denote unit vectors.

Focussing on the case $d=2$, the polarity vector can be expressed in polar coordinates and expanded around its average value,
\begin{equation}
	\bp(\tilde \br) =(p_0+\delta p(\tilde \br)) \begin{pmatrix}
		\cos(\theta(\tilde \br)) \\
		\sin(\theta(\tilde \br))
	\end{pmatrix}
	=
	(p_0+\delta p)  \bn \ .
\end{equation}
We now employ the spin-wave approximation, where the amplitude is considered fixed and not allowed to fluctuate, i.e., $\delta p=0$. 
One can then obtain an EOM for $\theta$ by projecting Eq.~\eqref{eq:eom} orthogonally to $\bn$, i.e., onto $\hat \bmm=(-\sin(\theta), \cos(\theta))^T$,
\begin{align}
\label{eq:eom_theta}
\nonumber
	\pp_t \theta = &- \lambda \cos(\theta) \pp_x \theta - \lambda \sin(\theta) \pp_y \theta + \mu_1 \nabla^2\theta + f \\
	&+  \mu_2 \Bigg[ \frac{1}{2}\cos(2\theta) ( \pp_y^2\theta - \pp_x^2\theta ) -\sin(2\theta) \pp_x\pp_y \theta \Bigg] + \mu_3 \Bigg[\frac{1}{2} \sin(2\theta)\Big((\pp_y \theta)^2-(\pp_x \theta)^2\Big) + \cos(2\theta) \pp_x\theta \pp_y \theta\Bigg]
\end{align}
where we have defined,
\begin{equation}
	\label{eq:theta}
	\lambda = \lambda_1 p_0 \sep \mu_1 = \mu_A+\frac{1}{2}(\mu_B+p_0^2\mu_C ) \sep \mu_2 = \mu_B - p_0^2\mu_C \sep \mu_3 = p_0^2(\mu_J+\mu_K+\mu_L-\mu_F) -\mu_B \ ,
\end{equation}
as well as $f$, with,
\begin{equation}
	\langle f(\tilde \br) f(\tilde \br^\prime) \rangle = 2D \, \delta^{3}(\tilde \br-\tilde \br^\prime)\ ,
\end{equation}
which is straightforwardly shown to be equivalent to $ \bmm \cdot \bff_p$. Since the couplings $\lambda$ and $\mu_1$, $\mu_2$ and $\mu_3$ are independent of $\theta$ due to rotational invariance, and since $\delta p=0$ due to the spin-wave approximation, there are only 4 independent parameters to characterize the completely generic EOM of Malthusian flocks to all orders in $\theta$ and second order in derivatives.

Note that if instead of imposing $\delta p=0$, one instead solves the EOM for $\bp$ in terms of $\theta$ by projecting Eq.~\eqref{eq:eom} onto $\bn$, and then inserting this solution into the EOM for $\theta$, retaining only terms to the same order in gradients, one obtaines the same EOM as Eq.~\eqref{eq:theta}, albeit with shifted coefficients. The universal physics does not depend on which path was chosen.

Compared to Ref.~\cite{chate_prl24,chen_a25}, where the nonlinear terms of second order in derivatives have been truncated, Eq.~\eqref{eq:theta} is fully invariant under rotations,
\begin{equation} \label{rotinv}
	\theta (\tilde \br)\to\theta(\tilde \br)+\psi, \, x\to x\cos\psi+y\sin\psi, \, y\to y\cos\psi-x\sin\psi \ .
\end{equation}

\section{Field Theoretic Formalism}

To apply the nonperturbative renormalization group (NPRG), the EOM \eqref{eq:theta} must first be converted to an action using the Martin-Siggia-Rose-de Dominicis-Janssen formalism \cite{martin_pra73,dedominicis_jpc76,janssen_zpb76,canet_jopa11}. In particular, for additive noise as in Eq.~\eqref{eq:theta} this is straightforward and yields,
\begin{align}
	\label{eq:action}
	S[\bar \theta,\theta] = \int_{\tilde \br} \Bigg\{&-D\bar\theta^2 +\bar \theta \Bigg(\partial_t\theta+ \lambda \cos(\theta) \pp_x \theta + \lambda \sin(\theta) \pp_y \theta - \mu_1 \nabla^2\theta \\
	\nonumber
	&-  \mu_2 \Big[\cos(2\theta) ( \pp_y^2\theta - \pp_x^2\theta ) -2\sin(2\theta) \pp_x\pp_y \theta \Big] - \mu_3 \Big[ \sin(2\theta)\Big((\pp_y \theta)^2-(\pp_x \theta)^2\Big) + 2\cos(2\theta) \pp_x\theta \pp_y \theta\Big] \Bigg) \Bigg\},
\end{align}
where $\int_{\tilde \br} = \int \dd^{d}r\dd t$ and the noise field $f$ has been traded for the response field $\bar \theta$. This action now describes the probability distribution for field configurations and thus the partition for this nonequilibrium model can be written as,
\begin{equation}
	Z[\bar J, J] = \int \mathcal D \bar \theta \mathcal D \theta \,  e^{-S[\bar \theta,\theta] + \int_r (\bar J \bar \theta + J \theta)} \ .
\end{equation}
As in equilibrium systems, the full statistical information can be obtained by applying functional derivatives to $Z$ with respect to the source fields $\bar J$ and $J$, giving the correlation and response functions of the system. Most of the time it is however convenient to consider other generating functionals (equivalent to different thermodynamic potentials in equilibrium) such as,
\begin{equation}
	W[\bar J,J] = \log Z \ ,
\end{equation}
which generates the connected correlation and response functions. For the NPRG formalism, the potential of interest is the effective average action,
\begin{equation}
	\Gamma[\bar \phi,\phi] = -W + \int_{\tilde \br}(\bar J \bar \phi+J \phi) \ ,
\end{equation}
i.e., the Legendre transformation of $W$ with respect to the source $J$, such that $\Gamma$ is a functional of the expectation value of the field $\phi = \delta W/\delta J = \langle \theta \rangle$ and the response field $\bar\phi = \delta W/\delta \bar J = \langle \bar\theta \rangle$.

In particular, the first order derivative of $\Gamma$ thus defines per definition an equation of motion for the expectation value of the fields,
\begin{equation}
	\frac{\delta \Gamma}{\delta \phi(\tilde \br)} = J(\tilde \br) \ ,
\end{equation}
and similarly for $\bar \phi$. Since $\Gamma$ is the Legendre transformation of $W$, it still contains all statistical information of the problem and all connected and disconnected correlation functions can be reconstructed from derivatives of $\Gamma$. Therefore, if $\Gamma$ is known, the many-body problem can be considered as solved.

If the action $S$ is quadratic, i.e., a free theory, then $\Gamma[\bar\phi,\phi]=S[\bar\phi,\phi]$ which is used as a vital ingredient to construct the exact flow equation discussed in the next section. In the following, we no longer distinguish between $\theta$ and $\phi$ or $\bar\theta$ and $\bar\phi$, as they are simply understood as the arguments of $\Gamma$.

\section{Nonperturbative Renormalization Group and Ansatz}

The functional renormalization group is implemented by modifying the action with a quadratic regulator term
\begin{equation}
	S[\bar\theta,\theta]\rightarrow S[\bar\theta,\theta]+\Delta S_k[\bar\theta,\theta] \ ,
\end{equation}
dependent on an arbitrary inverse length scale $k$. This modification of the theory also deforms the effective average action
\begin{equation}
    \Gamma[\bar\theta,\theta] \rightarrow \tilde\Gamma_k[\bar\theta,\theta] \ .
\end{equation}
Requiring that $\lim_{k\to \Lambda}\Delta S_k\gg S$, the modified action is dominated by the quadratic regulator part in the ultraviolet limit, such that $\tilde\Gamma_\Lambda= S+\Delta S_\Lambda$. Subtracting $\Delta S_k$ from $\tilde \Gamma$, thus creates the scale dependent effective action,
\begin{equation}
    \Gamma_k = \tilde \Gamma_k-\Delta S_k \ ,
\end{equation}
with the initial condition $\Gamma_\Lambda=S$. If on the other hand $\Delta S_0=0$, the original, nonmodified theory is restored. 

Through this construction, it is now straightforward to derive the exact functional RG equation, the Wetterich equation \cite{wetterich_plb93, morris_ijopa94,ellwanger_zfpc94},
\begin{equation}
	\label{eq:wetterich}
	\pp_k \Gamma_k =\frac{1}{2} {\rm Tr} \left[ \left(\Gamma^{(2)}_k +R_k\right)^{-1} \pp_k R_k\right]\ ,
\end{equation}
where $\Gamma_k^{(2)}$ is the Hessian of $\Gamma_k$, and $R_k$ is the Hessian of $\Delta S_k$ which is field independent since $\Delta S_k$ is quadratic. It can also be brought into the convenient form,
\begin{equation}
	\label{eq:wetterich2}
	\pp_k \Gamma_k =\frac{1}{2} \pp_{k'} {\rm Tr} \log \left(\Gamma^{(2)}_k +R_{k'}\right)\Bigg|_{k'=k} \ ,
\end{equation}
where $k'$ is taken independent from $k$ before taking the scale derivative.

\vspace{0.5cm}

The Wetterich equation can in general not be solved exactly. To define the nonperturbative RG approach, one needs to define an Ansatz for the flowing effective action and specify a regulator $\Delta S_k$. We choose the Ansatz to be of the same form as in the microscopic action \eqref{eq:action}, 
\begin{align}
	\label{eq:ansatz}
	\Gamma_k[\bar \theta,\theta] = \int_{\tilde \br} \Bigg\{&-D_k\bar\theta^2 +\bar \theta \Bigg(\gamma_k\partial_t\theta+ \lambda_k \cos(\theta) \pp_x \theta + \lambda_k \sin(\theta) \pp_y \theta - \mu_{1,k} \nabla^2\theta \\
	\nonumber
	&-  \mu_{2,k} \Big[\cos(2\theta) ( \pp_y^2\theta - \pp_x^2\theta ) -2\sin(2\theta) \pp_x\pp_y \theta \Big] - \mu_{3,k} \Big[ \sin(2\theta)\Big((\pp_y \theta)^2-(\pp_x \theta)^2\Big) + 2\cos(2\theta) \pp_x\theta \pp_y \theta\Big] \Bigg) \Bigg\},
\end{align}
albeit with scale dependent couplings, as we have already established above, that within the spin-wave approximation this form contains all terms allowed by symmetry at the second order of the derivative expansion on the level of the EOM for $\theta$. For completeness, we have also introduced a scale dependent coefficient of the time derivative term, even though it is not renormalized. While the noise term could take a more general form, here it is kept at leading order as a simple constant quadratic term as higher-order terms are not expected to be relevant. The choice of regulator is discussed in the next section.  

\section{Regulator Choice}

In principle with the requirements defined for the regulator in the previous section, for $k=\Lambda$ and $k=0$, the start and end-points of the RG flow are independent of the precise form of the regulator. However, by truncating the form of $\Gamma_k$ through the choice of an Ansatz, this property is typically broken and the results develop some form of regulator dependence. By comparing the results between different regulators, one can evaluate this residual dependence \cite{balog_pre20}.

In addition, in nonequilibrium statistical physics, one needs to pay attention that causality is respected by $\Delta S_k$ \cite{canet_jopa11}. The latter can straightforwardly be implemented by the form,
\begin{equation}
	\Delta S_k[\bar\theta,\theta] = \int_{\tilde \bq} \bar \theta(\tilde \bq) Q_k(\bq) \theta(\tilde \bq) \ ,
\end{equation}
where $\tilde \bq=(\omega_q,\bq)$ are frequencies and wave-vectors in Fourier-space and $\int_{\tilde \bq} = \int \dd^{d}q\dd \omega_q/(2\pi)^{d+1}$. In particular, the noise term, proportional to $\bar\theta^2$ is not regularized, and the regulator function $Q_k$ is independent of frequencies, which is indicated by the dependence on $\bq$ rather than $\tilde\bq$. 

Here we choose the $\bq$ independent form,
\begin{equation}
	Q_k(q) = (\mu_{1,k}+\mu_{2,k}) k^2 = \mu_{y,k} k^2  \ ,
\end{equation}
which is simply a scale dependent mass term which allows for analytic evaluation of the flow equations. While this regulator seemingly breaks the rotational invariance of the theory, it is easy to see that $\Gamma_k$ remains invariant throughout the RG flow as the breaking of the symmetry only happens at the linear level. 

Another form that respects causality is,
\begin{equation}
	\Delta S_k[\bar\theta,\theta] = \int_{\tilde \bq} \bar \theta(\tilde \bq) D_k(\bq) \bar \theta(\tilde \bq) \ ,
\end{equation}
which is also frequency independent, but modifies the noise term rather than the deterministic contribution. In particular, by setting,
\begin{equation}
\label{eq:cutoff_sharp}
	D_k(\bq) = D\big[\Theta(|\bq|-k) -1\big] \ ,
\end{equation}
one can straightforwardly implement a sharp cutoff realizing the Wilsonian momentum shell. The latter only works well when the noise is not renormalized as is the case here. 

\vspace{0.5cm}

With these two regulators, we can probe the regulator dependence of our results in two limiting cases, i.e., the flat (independent of $\bq$) and sharp (step function) limit, while at the same time being able to evaluate the momentum integrals analytically in both cases.

\section{Projections of Flow Equations}

The parameters appearing in the Ansatz \eqref{eq:ansatz} can be straightforwardly obtained from $\Gamma_k$ from the two and three-point functions, evaluated at an arbitrary constant background field $\theta(\tilde \bx)=\theta_0$ and vanishing response field $\bar\theta(\tilde \bx)=0$,
\begin{align}
\label{eq:f1}
    F_1(\theta_0,\tilde \bp) &= \frac{1}{VT} \frac{\delta^2\Gamma_k[0,\theta_0]}{\delta\theta(-\omega_p-\lambda_k/\gamma_k\bp\cdot\bn(\theta_0),-\bp)\delta\bar\theta(\omega_p+\lambda_k/\gamma_k\bp\cdot\bn(\theta_0),\bp)} \ , \\
    F_2(\theta_0,\tilde \bp) &= \frac{1}{VT} \frac{\delta^2\Gamma_k[0,\theta_0]}{\delta\bar\theta(-\omega_p-\lambda_k/\gamma_k\bp\cdot\bn(\theta_0),-\bp)\delta\bar\theta(\omega_p+\lambda_k/\gamma_k\bp\cdot\bn(\theta_0),\bp)} \ , \\
    F_3(\theta_0,\tilde \bp,\tilde \bh) &= \frac{1}{VT} \frac{\delta^3\Gamma_k[0,\theta_0]}{\delta\theta(\omega_h-\omega_p+\lambda_k/\gamma_k(\bh-\bp)\cdot\bn(\theta_0),\bh-\bp)\delta\theta(-\omega_h-\lambda_k/\gamma_k\bh\cdot\bn(\theta_0),-\bh)\delta\bar\theta(\omega_p+\lambda_k/\gamma_k\bp\cdot\bn(\theta_0),\bp)} ,
\end{align}
where $VT = (2\pi)^3\delta^{3}(0)$ is the spatiotemporal volume, $\bn(\theta_0) = (\cos(\theta_0),\sin(\theta_0))^T$ and 
where by shifting the external frequency, the expressions are evaluated in the co-moving frame.

From these functional derivatives, the couplings can then be obtained via,
\begin{align}
    \gamma_k &= -\ii \frac{\pp}{\pp \omega_p} F_1(0,\tilde \bp)\Big|_{\tilde \bp=0} \ , \\
    \lambda_k &= \ii \frac{\pp}{\pp p_x} F_1(0,\tilde \bp)\Big|_{\tilde \bp=0} \ , \\
    \mu_{x,k} &= \frac{1}{2}\frac{\pp^2}{\pp p_x^2} F_1(0,\tilde \bp)\Big|_{\tilde \bp=0} \ , \\
    \mu_{y,k} &= \frac{1}{2}\frac{\pp^2}{\pp p_y^2} F_1(0,\tilde \bp)\Big|_{\tilde \bp=0} \ , \\
    D_k &= -\frac{1}{2} F_2(0,\tilde \bp)\Big|_{\tilde \bp=0} \ , \\
    \label{eq:proj_mu3}
    \mu_3 &= \frac{1}{2}\frac{\pp^2}{\pp p_y\pp p_x} F_3(0,0,p_x,p_y,0,p_x,0)\Big|_{ \bp=0} \ , 
\end{align}
where we have introduced the linear combinations $\mu_{x,k}=\mu_{1,k}-\mu_{2,k}$ and $\mu_{y,k} = \mu_{1,k}+\mu_{2,k}$. In the following, we omit to write the $k$-dependence of the couplings to ease notation.
Applying the same projections to the Wetterich equation \eqref{eq:wetterich2} one obtains the flow equations of these couplings.

We have explicitly checked, that the projection of $\lambda$ does not depend on whether it is obtained from the 2-point or 3-point function, i.e, the projection 
\begin{equation}
    \lambda_k = \ii \frac{\pp}{\pp p_y}\frac{\pp}{\pp\theta_0} F_1(\theta_0,\tilde \bp)\Big|_{\tilde \bp=0,\theta_0=0} \ , 
\end{equation}
yields the same result for the flow equation of $\lambda$.

\section{Propagator and $n$-point functions}

The projections of the couplings from the previous section involve up to 3 functional derivatives. Since the right-hand-side of the Wetterich equation \eqref{eq:wetterich2} involves second order functional derivatives of $\Gamma_k$, applying the projections to Eq.~\eqref{eq:wetterich2} generates functional derivatives of $\Gamma_k$ up to fifth order, which are determined here.

We start with the regulated two-point function,
\begin{equation}
\label{eq:2point}
     \Gamma_k^{(2)}[0,\theta_0] + R_k = \bordermatrix{
    ~ & \bar \theta & \theta \cr
    \bar \theta & -2 D & \ii \gamma \omega_q + h(-\tilde \bq_1) \cr
    \theta & -\ii \gamma \omega_q + h(\tilde \bq_1) &  0
    } (2\pi)^3\delta^3(\tilde \bq_1+\tilde \bq_2) \ ,
\end{equation}
where we have defined,
\begin{equation}
\label{eq:freq_pole}
    h(\tilde\bq) = \ii \lambda \bq\cdot \bn(\theta_0) +\mu_1 (q_{x}^2+q_{y}^2)+\mu_2  \cos(2\theta_0) (q_{y}^2-q_{x}^2) - 2 \mu_2 q_x q_y \sin(2\theta_0) + Q_k(\bq) \ .
\end{equation}
We can thus straightforwardly invert Eq.~\eqref{eq:2point} to obtain the propagator,
\begin{equation}
\label{eq:gammat_i}
     G(\tilde \bq_1,\tilde \bq_2) =  (\Gamma_k^{(2)}[0,\theta_0] + R_k)^{-1} = \bordermatrix{
    ~ & \bar \theta & \theta \cr
    \bar \theta & 0 & \frac{1}{-\ii \gamma \omega_q+h(\tilde \bq_1)} \cr
    \theta & \frac{1}{\ii \gamma \omega_q+h(-\tilde \bq_1)} &  \frac{2D}{(-\ii \gamma \omega_q+h(\tilde \bq_1))(\ii \gamma \omega_q+h(-\tilde \bq_1))}
    } (2\pi)^3\delta^3(\tilde \bq_1+\tilde \bq_2) \ .
\end{equation}

The higher order $n$-point functions are presented, by keeping the full matrix structure of the first two functional derivatives and indicating with respect to which field the higher order derivatives were taken. For example $\Gamma_k^{(2;\bar\theta\theta)}$ implies that the Hessian of $\Gamma_k$ was derived first with respect to $\bar \theta$ and then with respect to $\theta$. The order of the derivatives matters for the labelling of the momenta.

We thus obtain,
\begin{align}
    \Gamma_k^{(2;\theta)}(\tilde \bq_1,\tilde \bq_2,\tilde \bq_3) &= \bordermatrix{
    ~ & \bar \theta & \theta \cr
    \bar \theta & 0 & \Gamma_k^{(\bar\theta\theta\theta)}(\tilde \bq_1,\tilde \bq_2,\tilde \bq_3) & \cr
    \theta & \Gamma_k^{(\theta\bar\theta\theta)}(\tilde \bq_1,\tilde \bq_2,\tilde \bq_3) &  0
    } \ , \\
    \Gamma_k^{(2;\bar\theta)}(\tilde \bq_1,\tilde \bq_2,\tilde \bq_3) &= \bordermatrix{
    ~ & \bar \theta & \theta \cr
    \bar \theta & 0 & 0 & \cr
    \theta & 0 &  \Gamma_k^{(\theta\theta\bar\theta)}(\tilde \bq_1,\tilde \bq_2,\tilde \bq_3)
    }  \ ,
    \\
    \Gamma_k^{(2;\theta\theta)}(\tilde \bq_1,\tilde \bq_2,\tilde \bq_3,\tilde \bq_4)  &= \bordermatrix{
    ~ & \bar \theta & \theta \cr
    \bar \theta & 0 & \Gamma_k^{(\bar\theta\theta\theta\theta)}(\tilde \bq_1,\tilde \bq_2,\tilde \bq_3,\tilde \bq_4) & \cr
    \theta & \Gamma_k^{(\theta\bar\theta\theta\theta)}(\tilde \bq_1,\tilde \bq_2,\tilde \bq_3,\tilde \bq_4) &  0
    }  \ ,
    \\
    \Gamma_k^{(2;\bar\theta\theta)}(\tilde \bq_1,\tilde \bq_2,\tilde \bq_3,\tilde \bq_4)  &= \bordermatrix{
    ~ & \bar \theta & \theta \cr
    \bar \theta & 0 & 0 & \cr
    \theta & 0 &  \Gamma_k^{(\theta\theta\bar\theta\theta)}(\tilde \bq_1,\tilde \bq_2,\tilde \bq_3,\tilde \bq_4)
    }  \ ,
    \\
    \Gamma_k^{(2;\bar\theta\bar\theta)}(\tilde \bq_1,\tilde \bq_2,\tilde \bq_3,\tilde \bq_4)  &=0\\
    \Gamma_k^{(2;\bar\theta\theta\theta)}(\tilde \bq_1,\tilde \bq_2,\tilde \bq_3,\tilde \bq_4,\tilde \bq_5)  &= \bordermatrix{
    ~ & \bar \theta & \theta \cr
    \bar \theta & 0 & 0 & \cr
    \theta & 0 &  \Gamma_k^{(\theta\theta\bar\theta\theta\theta)}(\tilde \bq_1,\tilde \bq_2,\tilde \bq_3,\tilde \bq_4,\tilde \bq_5)
    }  \ ,
\label{eq:gammat_f}
\end{align}
and the remaining $5$-point functions are not needed for the projections described in the previous section. All nonlinear terms involve exactly one derivative with respect to $\bar \theta$ as expected from the ansatz \eqref{eq:ansatz},
\begin{align}
\nonumber
\Gamma_k^{(\bar\theta\theta\theta)}(\tilde \bq_1,\tilde \bq_2,\tilde \bq_3) =(2\pi)^3&\delta^3(\tilde \bq_1+\tilde \bq_2+\tilde \bq_3)\\
\nonumber
    \times\Big\{ 
    \ \ &\ii \lambda \big[q_{1,x} \sin(\theta_0)-q_{1,y} \cos(\theta_0)\big]\\
\nonumber
    +& 2\mu_2 \big[\sin(2\theta_0)(q_{2,x}^2-q_{2,y}^2+q_{3,x}^2-q_{3,y}^2)-2\cos(2\theta_0)(q_{2,x}q_{2,y}+q_{3,x}q_{3,y})\big] \\
    -&2\mu_3\big[\sin(2\theta_0)(q_{2,x}q_{3,x}-q_{2,y}q_{3,y})-\cos(2\theta_0)(q_{2,x}q_{3,y}+q_{3,x}q_{2,y})\big]
    \Big\} \ ,
\end{align}
\begin{align}
\nonumber
\Gamma_k^{(\bar\theta\theta\theta\theta)}(\tilde \bq_1,\tilde \bq_2,\tilde \bq_3,\tilde \bq_4) =(2\pi)^3&\delta^3(\tilde \bq_1+\tilde \bq_2+\tilde \bq_3+\tilde \bq_4)\\
\nonumber
    \times\Big\{ 
    \ \ &\ii \lambda \big[q_{1,x} \cos(\theta_0) + q_{1,y} \sin(\theta_0)\big]\\
\nonumber
    +& 4\mu_2 \big[\cos(2\theta_0)(q_{2,x}^2-q_{2,y}^2+q_{3,x}^2-q_{3,y}^2+q_{4,x}^2-q_{4,y}^2)\\
    \nonumber
    &+2\sin(2\theta_0)(q_{2,x}q_{2,y}+q_{3,x}q_{3,y}+q_{4,x}q_{4,y})\big] \\
    \nonumber
    -& 4\mu_3 \big[ \cos(2\theta_0)(q_{2,x}q_{3,x}-q_{2,y}q_{3,y}+q_{2,x}q_{4,x}-q_{2,y}q_{4,y}+q_{3,x}q_{4,x}-q_{3,y}q_{4,y})\\
    &+\sin(2\theta_0)(q_{2,x}q_{3,y}+q_{2,y}q_{3,x}+q_{2,x}q_{4,y}+q_{2,y}q_{4,x}+q_{3,x}q_{4,y}+q_{3,y}q_{4,x})\big] 
    \Big\} \ ,
\end{align}
\begin{align}
\nonumber
\Gamma_k^{(\bar\theta\theta\theta\theta\theta)}(\tilde \bq_1,\tilde \bq_2,\tilde \bq_3,\tilde \bq_4,\tilde\bq_5) =(2\pi)^3&\delta^3(\tilde \bq_1+\tilde \bq_2+\tilde \bq_3+\tilde \bq_4+\tilde\bq_5)\\
\nonumber
    \times\Big\{ 
    \ \ &\ii \lambda \big[-q_{1,x} \sin(\theta_0) + q_{1,y} \cos(\theta_0)\big]\\
\nonumber
    -& 8\mu_2 \big[\sin(2\theta_0)(q_{2,x}^2-q_{2,y}^2+q_{3,x}^2-q_{3,y}^2+q_{4,x}^2-q_{4,y}^2+q_{5,x}^2-q_{5,y}^2)\\
    \nonumber
    &-2\cos(2\theta_0)(q_{2,x}q_{2,y}+q_{3,x}q_{3,y}+q_{4,x}q_{4,y}+q_{5,x}q_{5,y})\big] \\
    \nonumber
    +& 8\mu_3 \big[ \sin(2\theta_0)(q_{2,x}q_{3,x}-q_{2,y}q_{3,y}+q_{2,x}q_{4,x}-q_{2,y}q_{4,y}+q_{3,x}q_{4,x}-q_{3,y}q_{4,y}\\
    \nonumber
    &\ \ \ \  +q_{2,x}q_{5,x}-q_{2,y}q_{5,y}+q_{3,x}q_{5,x}-q_{3,y}q_{5,y}+q_{4,x}q_{5,x}-q_{4,y}q_{5,y})\\
    \nonumber
    &-\cos(2\theta_0)(q_{2,x}q_{3,y}+q_{2,y}q_{3,x}+q_{2,x}q_{4,y}+q_{2,y}q_{4,x}+q_{3,x}q_{4,y}+q_{3,y}q_{4,x}\\
    &\ \ \ \  +q_{2,x}q_{5,y}+q_{2,y}q_{5,x}+q_{3,x}q_{5,y}+q_{3,y}q_{5,x}+q_{4,x}q_{5,y}+q_{4,y}q_{5,x})\big] 
    \Big\} \ .
\end{align}

\section{Flow Equations}

By introducing the graphical notation,
\begin{align}
    G(\tilde \bq_1,\tilde\bq_2) = \diagram{10} \ , \\
    \Gamma^{(2,\bar\theta)}(\tilde \bq_1,\tilde\bq_2,\tilde\bq_3)= \diagram{11} \ , \\
    \Gamma^{(2,\theta)}(\tilde \bq_1,\tilde\bq_2,\tilde\bq_3)= \diagram{12} \ , \\
    \Gamma^{(2,\bar\theta\theta)}(\tilde \bq_1,\tilde\bq_2,\tilde\bq_3,\tilde\bq_4)= \diagram{13} \ , \\
    \Gamma^{(2,\theta\theta)}(\tilde \bq_1,\tilde\bq_2,\tilde\bq_3,\tilde\bq_4)= \diagram{14} \ , \\
    \Gamma^{(2,\bar\theta\theta\theta)}(\tilde \bq_1,\tilde\bq_2,\tilde\bq_3,\tilde\bq_4)= \diagram{15} \ , 
\end{align}
we can represent the flow equations of the two- and three-point functions, diagrammatically as,
\begin{align}
\label{eq:flow1}
    \partial_k \Gamma^{(\bar\theta\theta)}(\tilde \bq_1,\tilde\bq_2) &=  \frac{1}{2}\partial_{k'}\Bigg[-\diagram{2} +\diagram{3}\Bigg]_{k'=k} \ , \\
\label{eq:flow2}
    \nonumber
    \partial_k \Gamma^{(\bar\theta\theta\theta)}(\tilde \bq_1,\tilde\bq_2,\tilde \bq_3) &=  \frac{1}{2}\partial_{k'}\Bigg[\diagram{4} -\diagram{5}-\diagram{6}\\
    &\hspace{1.5cm}+2\diagram{7}-\diagram{8}\Bigg]_{k'=k} \ , 
\end{align}
where a contraction over connected elements is implied, realizing the trace of field indices, momenta and frequencies. The contractions over momenta and frequencies enforces ``conservation of momentum and energy", i.e., all momenta and frequencies of a vertex need to sum up to zero and the propagator conserves wavevector and frequencies also. One final loop integral over the remaining frequency and momentum remains. The contractions over field indices are realized as ordinary matrix multiplications of the matrices defined in Eqns.~\eqref{eq:gammat_i}-\eqref{eq:gammat_f}.

From the flow equations for the two- and three-point functions \eqref{eq:flow1} and \eqref{eq:flow2} the flow equations for the couplings can straightforwardly be deduced via \eqref{eq:f1}-\eqref{eq:proj_mu3}. The algebra is performed in the supplemental Mathematica notebook {\it ``Malthusian\_NPRG\_Smooth\_Regulator.nb"}. It involves the following steps:
\begin{enumerate}
    \item Multiplication of the Matrices \eqref{eq:gammat_i}-\eqref{eq:gammat_f} and integration over contracted momenta to obtain expressions for the diagrams \eqref{eq:flow1} and \eqref{eq:flow2}
    \item Projection of the flow equations for full two- and three-point functions onto the couplings via \eqref{eq:f1}-\eqref{eq:proj_mu3}
    \item Carrying out the frequency integral
    \item Carrying out the scale derivative
    \item Carrying out the momentum integral
\end{enumerate}

Applying all these steps yields the flow equations for the couplings,
\begin{align}
	\pp_\ell D &=\pp_l \gamma = 0 \ ,\\
	\pp_\ell \lambda &=- \frac{D (2-\eta_y) }{16\pi \gamma\sqrt{\mu_x\mu_y^3}}\lambda\Big(\mu_x+\mu_y-\mu_3\Big) \ , \\
	\pp_\ell \mu_x &= \frac{D (2-\eta_y) }{16\pi \gamma\sqrt{\mu_x\mu_y^3}} \Big(2\mu_3^2-3\mu_3(\mu_x-\mu_y)+\mu_x^2-6 \mu_x\mu_y+5\mu_y^2\Big) \ , \\
	\pp_\ell \mu_y &= \frac{D (2-\eta_y) }{16\pi \gamma\sqrt{\mu_x\mu_y^3}} \Bigg(\frac{3}{2}\frac{\lambda^2}{k^2} +\frac{\mu_y}{\mu_x} \Big(2\mu_3^2-3\mu_3(\mu_x-\mu_y)+5 \mu_x^2-6 \mu_x\mu_y+ \mu_y^2\Big) \Bigg) \ , \\
	\pp_\ell \mu_3 &= \frac{D (2-\eta_y)}{16\pi \gamma\sqrt{\mu_x\mu_y^3}} \Bigg(\frac{\lambda^2}{24k^2\mu_y}\Big(13\mu_3-13\mu_x-23\mu_y\Big) -8\mu_3\mu_y\Bigg)\ ,
\end{align}
where we have defined the scale parameter $\ell=-\log k/\Lambda$, often referred to as ``RG-time".

To obtain fixed point solutions, the flow equations need to be brought into dimensionless form, by rescaling the couplings,
\begin{align}
\label{eq:dimensionless_lambda}
	\bar \lambda = \sqrt\frac{D}{{4\pi\gamma k^2}\sqrt{\mu_x\mu_y^5}} \lambda \ , \\
    \label{eq:dimensionless_mux}
	\bar \mu_x = \frac{D}{4\pi \gamma} \frac{1}{\sqrt{\mu_x^3 \mu_y}} \mu_x \ , \\
    \label{eq:dimensionless_muy}
	\bar \mu_y = \frac{D}{4\pi \gamma} \frac{1}{\sqrt{\mu_x^3 \mu_y}} \mu_y \ , \\
    \label{eq:dimensionless_mu3}
	\bar \mu_3 = \frac{D}{4\pi \gamma} \frac{1}{\sqrt{\mu_x^3 \mu_y}} \mu_3 \ . 
\end{align}
The factor $1/4\pi$ was introduced for convenience. We also introduce the anomalous dimensions,
\begin{align}
\label{eq:defetax}
    \eta_x = \pp_\ell \mu_x \ , \\
\label{eq:defetay}
    \eta_y = \pp_\ell \mu_y \ ,
\end{align}
whose fixed point value defines the anomalous scaling exponents. While in the presence of a sharp regulator, these suffice as definitions for $\eta_x$ and $\eta_y$, for a smooth regulator the right-hand-side of Eqns.~\eqref{eq:defetax} and \eqref{eq:defetay} depends also $\eta_y$. Eq.~\eqref{eq:defetay} thus defines an algebraic equation for $\eta_y$ which can be solved straightforwardly,
\begin{align}
	\eta_x &=(2-\eta_y) \frac{2\bar\mu_3^2-3\bar\mu_3(\bar\mu_x-\bar\mu_y)+\bar\mu_x^2-6\bar\mu_x\bar\mu_y+5\bar\mu_y^2}{4\bar\mu_y} \ , \\
	\eta_y &= \frac{2 \left(3 \bar\lambda ^2 \bar\mu_y+4 \bar\mu_3^2 -6 \bar\mu_3 \bar\mu_x+6 \bar\mu_3
   \bar\mu_y+10 \bar\mu_x^2 +2  \bar\mu_y^2-12 \bar\mu_x \bar\mu_y\right)}{3 \bar\lambda ^2 \bar\mu_y+4 \bar\mu_3^2 -6 \bar\mu_3 \bar\mu_x+6 \bar\mu_3
   \bar\mu_y+10 \bar\mu_x^2 +2  \bar\mu_y^2-12 \bar\mu_x \bar\mu_y+8  \bar\mu_y} \ . 
\end{align}
This creates in particular a nonpolynomial dependence of the flow equations on the couplings, a manifestation of nonperturbative effects \cite{canet_a09,jentsch_a24}.

Note that since $\mu_x$ and $\mu_y$ contribute both linear and nonlinear terms to the EOM, which are tied together by rotational symmetry, $\mu_x$ and $\mu_y$ do not completely disappear in the flow equations as is usually the case. Instead, a dimensionless rescaled version of these couplings, \eqref{eq:dimensionless_mux} and \eqref{eq:dimensionless_muy}, remain:
\begin{align}
	\pp_l \bar \lambda  &= (1-\frac{1}{4}\eta_x-\frac{5}{4}\eta_y)\bar \lambda +  \frac{2-\eta_y}{4}\frac{\bar\mu_3-\bar\mu_x-\bar\mu_y}{\bar \mu_y} \bar \mu_x \bar \lambda \ , \\
	\pp_l \bar \mu_x  &= \frac{1}{2}(-3\eta_x-\eta_y)\bar \mu_x +  \eta_x \bar \mu_x \ , \\
	\pp_l \bar \mu_y  &= \frac{1}{2}(-3\eta_x-\eta_y)\bar \mu_y +  \eta_y \bar \mu_y \ , \\
    \label{eq:dimless_flow_mu3}
	\pp_l \bar \mu_3  &= \frac{1}{2}(-3\eta_x-\eta_y)\bar \mu_3 + \frac{2-\eta_y}{96} \Big(\bar \lambda^2(13\bar\mu_3-13\bar \mu_x-23\bar\mu_y)-192\bar\mu_3\bar\mu_x\Big) \ .
\end{align}

These flow equations can straightforwardly be solved numerically, and the locations of the fixed points can even be determined analytically. The results are reported in the main text and details of the calculation are given in the supplemental Mathematica notebook {\it ``Malthusian\_NPRG\_Smooth\_Regulator.nb"}.

\vspace{0.5cm}

The two-dimensional sections of this four-dimensional coupling space, presented in Fig.~1 of the main text, are obtained in the following way. For Fig.~1a, setting $\bar\mu_x=0$ immediately identifies the attractive manifold that contains both the stable Malthusian fixed point and Toner's unstable fixed point. To further reduce the dimensionality, of the RG flow, we assume that the nonlinear coupling $\bar\mu_3$ has converged to its fixed point value given the current value of the other couplings. We thus set $\bar\mu_3$ to the value given by the stationary solution of Eq.~\eqref{eq:dimless_flow_mu3}. The second slice in Fig.~1b, highlights the RG flow near the free theory. Since $\bar\lambda$ governs the stability of the free theory, we set the remaining nonlinearities to zero, i.e., $\mu_3=0$ and $\mu_2=(\mu_y-\mu_x)/2=0$.

\section{Flow equations with sharp regulator}

To probe the dependence of our results on the choice of regulator, we also compute the flow equations in the presence of a sharp cutoff \eqref{eq:cutoff_sharp}. For this regulator, the evaluation of the flow equations differs slightly.

\begin{enumerate}
    \item Multiplication of the Matrices \eqref{eq:gammat_i}-\eqref{eq:gammat_f} and integration over contracted momenta to obtain expressions for the diagrams \eqref{eq:flow1} and \eqref{eq:flow2}
    \item Projection of the flow equations for full two- and three-point functions onto the couplings via \eqref{eq:f1}-\eqref{eq:proj_mu3}
    \item The term $Q_k$ in Eq.~\eqref{eq:freq_pole} is set to zero
    \item The frequency integral is carried out
    \item $D$ is replaced via $D\rightarrow D\,\Theta(q_y-k)$
    \item The scale derivative turns the Heaviside function in to $k\delta(q_y-k)$, enabling straightforward evaluation of the $q_y$ integral.
    \item The integration over $q_x$ is carried out.
\end{enumerate}

The resulting flow equations are,
\begin{align}
	\pp_l D &=\pp_l \gamma = 0 \ ,\\
    \label{eq:flow_lambda_sharp}
	\pp_l \lambda &= -\frac{D }{8\pi \gamma\sqrt{\mu_x\mu_y^3}}\lambda\Big(\mu_x+\mu_y-\mu_3\Big) \ , \\
	\pp_l \mu_x &= \frac{D }{8\pi \gamma\sqrt{\mu_x\mu_y^3}} \Big(2\mu_3^2-3\mu_3(\mu_x-\mu_y)+\mu_x^2-6 \mu_x\mu_y+5\mu_y^2\Big) \ , \\
	\pp_l \mu_y &= \frac{D }{8\pi \gamma\sqrt{\mu_x\mu_y^3}} \Bigg(\frac{5}{4}\frac{\lambda^2}{k^2} +\frac{\mu_y}{\mu_x} \Big(2\mu_3^2-3\mu_3(\mu_x-\mu_y)+5 \mu_x^2-6 \mu_x\mu_y+ \mu_y^2\Big) \Bigg) \ , \\
	\pp_l \mu_3 &= \frac{D }{8\pi \gamma\sqrt{\mu_x\mu_y^3}} \Bigg(\frac{\lambda^2}{16k^2\mu_y}\Big(23\mu_3-23\mu_x+7\mu_y\Big) -8\mu_3\mu_y\Bigg)\ .
\end{align}

Using the same rescalings as above \eqref{eq:dimensionless_lambda}-\eqref{eq:dimensionless_mu3}, the flow equations can similarly be brought into dimensionless form,
\begin{align}
	\pp_l \bar \lambda  &= (1-\frac{1}{4}\eta_x-\frac{5}{4}\eta_y)\bar \lambda +  \frac{1}{2}\frac{\bar\mu_3-\bar\mu_x-\bar\mu_y}{\bar \mu_y} \bar \mu_x \bar \lambda \ , \\
	\pp_l \bar \mu_x  &= \frac{1}{2}(-3\eta_x-\eta_y)\bar \mu_x +  \eta_x \bar \mu_x \ , \\
	\pp_l \bar \mu_y  &= \frac{1}{2}(-3\eta_x-\eta_y)\bar \mu_y +  \eta_y \bar \mu_y \ , \\
	\pp_l \bar \mu_3  &= \frac{1}{2}(-3\eta_x-\eta_y)\bar \mu_y + \frac{1}{32} \Big(\bar \lambda^2(23\bar\mu_3-23\bar \mu_x+7\bar\mu_y)-128\bar\mu_3\bar\mu_x\Big) \ , 
\end{align}
where the anomalous dimensions now have a simpler form,
\begin{align}
	\eta_x &= \frac{2\bar\mu_3^2-3\bar\mu_3(\bar\mu_x-\bar\mu_y)+\bar\mu_x^2-6\bar\mu_x\bar\mu_y+5\bar\mu_y^2}{2\bar\mu_y} \ , \\
	\eta_y &= \frac{8\bar\mu_3^2-12\bar\mu_3(\bar\mu_x-\bar\mu_y)+20\bar\mu_x^2-24\bar\mu_x\bar\mu_y+4\bar\mu_y^2+5\bar\lambda^2\bar\mu_y}{8\bar\mu_y} \ . \\
\end{align}
The calculations arriving at these flow equations are presented in {\it ``Malthusian\_NPRG\_Sharp\_Regulator.nb"}.

The qualitative features of the RG flow, i.e., the fixed point structure is not changed with this different choice of regulator, nor are the quantitative values of the scaling exponents, demonstrating the robustness of our approach. What changes instead is the precise location of the fixed points. With the sharp regulator we obtain the following location for the Malthusian fixed point which needs to be obtained numerically,
\begin{equation}
    \lambda^*\approx 0.95 \ , \ \ \ \ \mu_x^* = 0 \ , \ \ \ \ \mu_y^* \approx 0.03 \ , \ \ \ \ \mu_3^* \approx 0.05 \ .
\end{equation}
Note that $\mu_3^*$ has opposite sign compared to the fixed point value with the smooth regulator.

For Toner's fixed point, we obtain,
\begin{equation}
    \lambda^*\approx \frac{4\sqrt2}{5} \ , \ \ \ \ \mu_x^* = 0 \ , \ \ \ \ \mu_y^* = 0 \ , \ \ \ \ \mu_3^* = 0 \ .
\end{equation}
Interestingly, using this regulator, this fixed point acquires one additional unstable direction compared to the smooth regulator. The behaviour of the attractive line of fixed points, in particular also the location of the critical end-point is unchanged.

\section{Comparison with Dynamic Renormalization Group Formalism}

In the NPRG formalism described in the main text and above the different nonlinearities that would arise from expansion of the EOM into a power series of $\theta$ but are constrained together through rotational symmetry are characterized by a single coupling, i.e., $\lambda$, $\mu_2$ and $\mu_3$. These are thus described by a single flow equation, such that rotational invariance is manifest. This differs from the dynamic renormalization group approach taken in Refs.~\cite{chate_prl24,chen_a25} where the quadratic and cubic nonlinearities stemming from $\lambda$ are treated as a priori independent couplings, which happen to be equal initially and, due to the rotational symmetry realized by this equality, remain equal throughout the RG flow. As a consequence of this treatment, the two couplings can be rescaled independently.

To facilitate a comparison with this more common approach, we rederive the flow equations for $\lambda_x^{(1)}$, $\lambda_y^{(2)}$ and $\lambda_x^{(3)}$, the linear, quadratic and cubic contributions of the $\lambda$ nonlinearity, in the situation where the effective action is expanded into a power-series in $\theta$, and where all nonlinear terms arising through this are treated as independent, i.e., we use the Ansatz,
\begin{align}
	\label{eq:ansatz2}
	\nonumber
	\Gamma_k[\bar \theta,\theta] = \int_{\tilde \br} \Bigg\{&-D\bar\theta^2 +\bar \theta \Bigg(\gamma\partial_t\theta+ \lambda_x^{(1)} \pp_x \theta - \frac{\lambda_x^{(3)}}{2} \theta^2 \pp_x \theta + \frac{\lambda_x^{(5)}}{24} \theta^4 \pp_x \theta +  \lambda_y^{(2)} \theta \pp_y \theta -\frac{\lambda_y^{(4)}}{6} \theta^3 \pp_y \theta - \mu_x \pp_x^2 \theta - \mu_y \pp_y^2 \theta \\
    \nonumber
	&+   2 \mu_{2}^{(3y)} \theta^2  \pp_y^2\theta - 2\mu_{2}^{(3x)} \theta^2 \pp_x^2\theta  +4\mu_{2}^{(2xy)}\theta \pp_x\pp_y \theta-  \frac{8}{3} \mu_{2}^{(4xy)} \theta^3 \pp_x\pp_y \theta \\
    & -2\mu_{3}^{(3y)}\theta(\pp_y \theta)^2 +2\mu_{3}^{(3x)}\theta(\pp_x \theta)^2 - 2\mu_{3}^{(2xy)} \pp_x\theta \pp_y \theta + 4\mu_{3}^{(4xy)}   \theta^2 \pp_x\theta \pp_y \theta \Bigg)  \Bigg\} \ .
\end{align}
We have expanded the Ansatz to fifth order in $\theta$ for the $\lambda$ nonlinearity, but only to fourth order for the nonlinearities stemming from $\mu_2$ and $\mu_3$. The reason is that, as we have seen above, the graphical corrections of $\lambda$ are proportional to $\lambda$ itself. Therefore, the contribution to the flow equation from the diagram involving the six-point vertex can only come from $\lambda$.

The DRG flow equations are obtained in a supplemental Mathematica notebook {\it ``Malthusian\_DRG\_Comparison.nb"},
\begin{align}
\pp_l \lambda_x^{(1)} = -\frac{D}{8\pi\gamma \sqrt{\mu_x\mu_y}} \Bigg(&2\lambda^{(3)}_x-\frac{2\lambda_y^{(2)}\mu_{2}^{(2xy)}}{\mu_y}-\frac{\lambda_y^{(2)}\mu_{3}^{(2xy)}}{\mu_y}\Bigg) \ ,
\\
\nonumber
\label{eq:drg_lam2}
\pp_l \lambda_y^{(2)} = -\frac{D}{8\pi\gamma \sqrt{\mu_x\mu_y}} \Bigg(&2 \lambda_y^{(4)} + \frac{10 \lambda_y^{(2)} \mu_{2}^{(3x)}}{\mu_{x}}-\frac{24 \lambda_y^{(2)} (\mu_{2}^{(2xy)})^2}{\mu_{x} \mu_{y}}-\frac{8 \lambda_y^{(2)} \mu_{2}^{(2xy)} \mu_{3}^{(2xy)}}{\mu_{x} \mu_{y}}-\frac{14 \lambda_y^{(2)} \mu_{2}^{(3y)}}{\mu_{y}}+\frac{3 \lambda_y^{(2)} \mu_{3}^{(3x)}}{\mu_{x}}\\
&-\frac{5 \lambda_y^{(2)} \mu_{3}^{(3y)}}{\mu_{y}}+\frac{2 \lambda_x^{(3)} \mu_{2}^{(2xy)}}{\mu_{x}}+\frac{\lambda_x^{(3)} \mu_{3}^{(2xy)}}{\mu_{x}} \Bigg) \ ,
\\
\nonumber
\label{eq:drg_lam3}
\pp_l \lambda_x^{(3)} = -\frac{D}{8\pi\gamma \sqrt{\mu_x\mu_y}} \Bigg(&2 \lambda_x^{(5)} + \frac{32\lambda_x^{(3)}\mu_{2}^{(3x)}}{\mu_x}+\frac{10\lambda_x^{(3)}\mu_{3}^{(3x)}}{\mu_x}+\frac{12\lambda_y^{(2)}\mu_{2}^{(2xy)}\mu_{2}^{(3y)}}{\mu_y^2} + \frac{6\lambda_y^{(2)}\mu_{2}^{(3y)}\mu_{3}^{(2xy)}}{\mu_y^2}+\frac{24\lambda_y^{(2)}(\mu_{2}^{(2xy)})^3}{\mu_x\mu_y^2}\\
\nonumber
&+\frac{12\lambda_y^{(2)}(\mu_{2}^{(2xy)})^2\mu_{3}^{(2xy)}}{\mu_x\mu_y^2}-\frac{2\lambda_y^{(4)}\mu_{2}^{(2xy)}}{\mu_y}-\frac{24\lambda_x^{(3)}\mu_{2}^{(3y)}}{\mu_y}-\frac{\lambda_y^{(4)}\mu_{3}^{(2xy)}}{\mu_y}-\frac{6\lambda_x^{(3)}\mu_{3}^{(3y)}}{\mu_y}-\frac{8\lambda_y^{(2)}\mu_{2}^{(4xy)}}{\mu_y}\\
\nonumber
&-\frac{4\lambda_y^{(2)}\mu_{3}^{(4xy)}}{\mu_y}-\frac{20\lambda_y^{(2)}\mu_{2}^{(3x)}\mu_{2}^{(2xy)}}{\mu_x\mu_y}-\frac{56\lambda_x^{(3)}(\mu_{2}^{(2xy)})^2}{\mu_x\mu_y}-\frac{8\lambda_y^{(2)}\mu_{2}^{(2xy)}\mu_{3}^{(3x)}}{\mu_x\mu_y}\\
&-\frac{2\lambda_y^{(2)}\mu_{2}^{(3x)}\mu_{3}^{(2xy)}}{\mu_x\mu_y}-\frac{16\lambda_x^{(3)}\mu_{2}^{(2xy)}\mu_{3}^{(2xy)}}{\mu_x\mu_y}\Bigg) \ .
\end{align}
Note first, that by setting 
\begin{align}
\label{eq:rest_sym1}
\lambda_x^{(1)}&=\lambda_y^{(2)}=\lambda_x^{(3)}=\lambda_y^{(4)}=\lambda_x^{(5)} \ , \\ \mu_{2}^{(3x)}&=\mu_{2}^{(3y)}=\mu_{2}^{(2xy)}=\mu_{2}^{(4xy)}=\mu_2 = (\mu_y-\mu_x)/2 \ ,\\
\label{eq:rest_sym3}
\mu_{3}^{(3x)}&=\mu_{3}^{(3y)}=\mu_{3}^{(2xy)}=\mu_{3}^{(4xy)}=\mu_3 \ ,
\end{align}
symmetry is restored and the flow equations coincide with each other and with Eq.~\eqref{eq:flow_lambda_sharp}.

We now eliminate irrelevant couplings. For this we first rescale,
\begin{equation}
    y\to y e^{\ell} \ ,\ \ \ \ x\to x e^{\zeta\ell}\ ,\ \ \ \ t\to t e^{z\ell}\ ,\ \ \ \ \theta\to \theta e^{\chi\ell} \ ,
\end{equation}
to obtain the DRG equations,
\begin{align}
    \label{drg:D}
    \pp_\ell D &= [z-1-\zeta-2\chi]D  \ ,\\
    \pp_\ell \mu_x &= [z-2\zeta +\eta_x]\mu_x  \ ,\\
    \pp_\ell \mu_y &= [z-2 +\eta_y]\mu_y  \ ,\\
    \pp_\ell \lambda^{(n)}_x &= [z-\zeta+(n-1)\chi +\eta_\lambda]\lambda_x^{(n)} && (n\ \ {\rm odd}) \ , \\
    \pp_\ell \lambda^{(n)}_y &= [z-1+(n-1)\chi +\eta_\lambda]\lambda_y^{(n)}  && (n\ \ {\rm even}) \ , \\
    \pp_\ell \mu_2^{(nx)} &= [z-2\zeta+(n-1)\chi +\eta_{\mu_2}]\mu_2^{(nx)} && (n\ \ {\rm odd}) \ ,\\
    \pp_\ell \mu_2^{(nxy)} &= [z-1-\zeta+(n-1)\chi +\eta_{\mu_2}]\mu_2^{(nxy)} && (n\ \ {\rm even}) \ ,\\
    \pp_\ell \mu_2^{(ny)} &= [z-2+(n-1)\chi +\eta_{\mu_2}]\mu_2^{(ny)} && (n\ \ {\rm odd}) \ ,\\
    \pp_\ell \mu_3^{(nx)} &= [z-2\zeta+(n-1)\chi +\eta_{\mu_3}]\mu_3^{(nx)} && (n\ \ {\rm odd}) \ ,\\
    \pp_\ell \mu_3^{(nxy)} &= [z-1-\zeta+(n-1)\chi +\eta_{\mu_3}]\mu_3^{(nxy)} && (n\ \ {\rm even}) \ ,\\
    \pp_\ell \mu_3^{(ny)} &= [z-2+(n-1)\chi +\eta_{\mu_3}]\mu_3^{(ny)} && (n\ \ {\rm odd}) \ ,\\
\end{align}
where, as in Ref.~\cite{chen_a25}, we have already set equal the graphical corrections $\eta$ of the nonlinearities, that are the same due to rotational symmetry.

As noise is not renormalized, there is no graphical correction for $D$ in Eq.~\eqref{drg:D}. At a fixed point, this imposes the hyperscaling relation,
\begin{equation}
    z-1-\zeta-2\chi=0 \ ,
\end{equation}
as discussed in Refs.~\cite{toner_prl12,chate_prl24,chen_a25}.

By evaluating these equations near Toner's fixed point \cite{toner_prl12}, both Refs.~\cite{chate_prl24} and \cite{chen_a25} realized, that in addition to $\lambda_y^{(2)}$ also $\lambda_x^{(3)}$ is a relevant nonlinearity driving the RG flow away from this fixed point.
In Ref.~\cite{chen_a25}, it was then observed that due to rotational symmetry, the graphical correction of both $\lambda_y^{(2)}$ and $\lambda_x^{(3)}$ must be the same. Since at a potential RG fixed point with nontrivial fixed point values of $\lambda_y^{(2)}$ and $\lambda_y^{(3)}$, the DRG equations imply
\begin{align}
    0&=z-\zeta+2\chi +\eta_\lambda \ , \\
    0&=z-1+\chi +\eta_\lambda \ ,
\end{align}
at such a fixed point. Subtracting the two equations from one another then implies a scaling relation,
\begin{equation}
    \label{eq:hyper2}
    1-\zeta+\chi = 0 \ ,
\end{equation}
which is independent of the actual value taken by $\eta_\lambda$ and thus exact. 

A similar line of argument can be used to show that the nonlinearities $\mu_2^{(2xy)}$ and $\mu_2^{(3x)}$ are also relavant at Toner's fixed point and must therefore be considered at the Malthusian fixed point. Evaluating the DRG equations for these couplings at Toner's fixed point \cite{toner_prl12}, $z=6/5$, $\zeta=3/5$ and $\chi=-1/5$, yields,
\begin{align}
    \pp_\ell \mu_2^{(3x)}&= [z-2\zeta+2\chi +\eta_{\mu_2}]\mu_2^{(3x)} = [-2/5 +\eta_{\mu_2}]\mu_2^{(3x)} ,\\
    \pp_\ell \mu_2^{(2xy)} &= [z-1-\zeta+\chi +\eta_{\mu_2}]\mu_2^{(2xy)} = [-3/5+\eta_{\mu_2}]\mu_2^{(2xy)}   \ .
\end{align}
Seemingly, the value of $\eta_{\mu_2}$ is a priori not known, and therefore one typically assumes it remains small enough to not change the relevance of the coupling. Using this argument one would conclude that $\mu_2^{(2xy)}$ and $\mu_2^{(3x)}$ are irrelevant. This is however incorrect since, due to rotational symmetry, the graphical correction of $\mu_2$ is linked to that of $\mu_x$ and $\mu_y$ via $\mu_2^{(1y)} = \mu_2^{(2xy)} = \mu_2^{(3x)} = \mu_2 = (\mu_y-\mu_x)/2$. Since $\eta_y=4/5>\eta_x=0$ at this fixed point, $\mu_2$ is dominated by the contribution from $\mu_y$ and therefore $\eta_{\mu_2}=\eta_y=4/5$. Using this result we straightforwardly see that both $\mu_2^{(2xy)}$ and $\mu_2^{(3x)}$ are relevant nonlinearities at Toner's fixed point. Since $\mu_3^{(2xy)}$ and $\mu_3^{(3x)}$ are of the same order in fields and derivatives, they should also be included in the analysis, even if the value of $\eta_{\mu_3}$ is not constrained by symmetry.

As one can straightforwardly check, using the exponents obtained at the Malthusian fixed point,
\begin{equation}
    \chi = -\frac{1}{4} \ , \ \ \ \zeta = \frac{3}{4} \ , \ \ \ z=\frac{5}{4} \ , \ \ \ \eta_x = \frac{1}{4} \ , \ \ \ \eta_y =\eta_{\mu_2} = \frac{3}{4} \ ,
\end{equation}
the DRG flow equations of all relevant nonlinearities, $\lambda_y^{(2)}$, $\lambda_x^{(3)}$, $\mu_2^{(2xy)}$ and $\mu_2^{(3x)}$ vanish. Since $\mu_3$ has a nontrivial fixed point value we have $\eta_{\mu_3}=\eta_y$, confirming a posteriori that $\mu_3^{(2xy)}$ and $\mu_3^{(3x)}$ cannot be discarded. All other nonlinear terms are irrelevant.

By eliminating all irrelevant couplings in Eqns.~\eqref{eq:drg_lam2} and \eqref{eq:drg_lam3} and restoring symmetry \eqref{eq:rest_sym1}-\eqref{eq:rest_sym3} in the limit $\mu_2\to \mu_y/2$ one can produce the perturbative one-loop result in the presence of all relevant nonlinear terms. This shows that, at the one-loop level, the graphical corrections of $\lambda_y^{(2)}$ and $\lambda_x^{(3)}$ do indeed vanish at the Malthusian fixed point, thus realizing the scaling exponents predicted in Ref.~\cite{chate_prl24}. Renormalization at the two-loop level is however not excluded, which likely implies that two-loop corrections of $\lambda$ are not captured by our second order derivative expansion.

\bibliography{references}